\documentclass[10pt,letterpaper]{article}

\usepackage[top=0.85in,left=2.75in,footskip=0.75in]{geometry}

\usepackage{amsmath,amssymb}

\usepackage{latexsym,bm}

\usepackage{changepage}

\usepackage[utf8x]{inputenc}

\usepackage{textcomp,marvosym}

\usepackage{cite}

\usepackage{nameref,hyperref}

\usepackage[right]{lineno}

\usepackage{microtype}
\DisableLigatures[f]{encoding = *, family = * }

\usepackage[table]{xcolor}

\usepackage{array}

\usepackage{color}

\usepackage{braket}

\newcolumntype{+}{!{\vrule width 2pt}}

\newlength\savedwidth

\raggedright
\usepackage[aboveskip=1pt,labelfont=bf,labelsep=period,justification=raggedright,singlelinecheck=off]{caption}

\makeatletter
\renewcommand{\@biblabel}[1]{\quad#1.}
\makeatother

\date{}

\usepackage{lastpage,fancyhdr,graphicx}
\usepackage{epstopdf}
\fancyheadoffset[L]{2.25in}
\fancyfootoffset[L]{2.25in}

\begin{document}
\vspace*{0.2in}

\begin{flushleft}
{\Large
\textbf\newline{\textbf{A new additive decomposition of velocity gradient}} 
}
\newline
\\
Bohua Sun\textsuperscript{\Yinyang}
\\
\bigskip
\textbf{\Yinyang}School of Civil Engineering \& Institute of Mechanics and Technology\\
Xi'an University of Architecture and Technology, Xi'an 710055, China
\\
\href{imt}{http://imt.xauat.edu.cn}
\bigskip

%
%





* sunbohua@xauat.edu.cn

\end{flushleft}
\section*{Abstract}
To avoid the infinitesimal rotation nature of the Cauchy-Stokes decomposition of velocity gradient, the letter proposes an new additive decomposition in which one part is a SO(3) rotation tensor $\bm Q=\exp \bm W$.

\section*{Author summary}
Bohua Sun (born 21 December 1963) is a Chinese scientist and an academician (Member) of the Academy of Science of South Africa. He was born in Tongshan in Xuzhou City.

His tenured full professorship at Cape Peninsula University of Technology (CPUT) started from 1995 and full professor since 2000, where he was the Director of the Center for Mechanics and Technology. From December 18,2018, he returned to China, and take a tenured Chair professorship at Xi'an University of Architecture and Technology, where he is Director and Chief Scientist of the Institute of Mechanics and Technology (IMT).

He graduated from Chang-An University in 1983 with B.Sc.(Eng), and received his M.Sc.(Eng) from Xi'an University of Architecture and Technology in 1986. In 1989, he received a Ph.D. in mechanics from Lanzhou University. Then he became one of the first Post-doc in the Dept. of Engineering Mechanics at Tsinghua University from 1989 to 1991.

From 1991, he studied abroad, from 1991-91 he was Research Fellow in the Faculty of Aerospace Engineering at Delft University of Technology, from 1992-93 an Alexander von Humboldt Research Fellow at Ruhr Universit\"at Bochum and from 1994 to 1995 a Post-doc Associate at Faculty of Engineering at University of Cape Town.

He is Editor-in-Chief of academic book series: Advances in Materials and Mechanics, Chief Editor of Advances in Engineering Mechanics, as well as founding editor of Journal of Nanomaterials.

In 2010 he was inducted into the Academy of Science of South Africa. He had been elected as Top 10 Overseas Chinese in the News of year 2010, and Top 10 Overseas Chinese in Washington Chinese Yearbook 2010, and awarded as outstanding post-doctoral alumina of Tsinghua University in 2018.

He has been awarded with Platinum Award for research and publication for 2016-2017 by the Cape Peninsula University of Technology.



\newpage

The vortex identification is a quite important tool for turbulence study \cite{jeong,cha,bos,tardu,elsas,epps,sun2019}. Recently, to have a better vortex identification, a vector named rortex vector was proposed \cite{liu-2,liu-3,liu-4,liu-5,liu-6,liu-7}. The basic arguments is that the vorticity $\bm \omega=\bm \nabla \times \bm v$ would not able to represent vortex (rotation) in fluid, should be further decomposed into two parts, namely
\begin{equation}\label{liu}
  \bm \omega=\bm \nabla \times \bm v= \bm R+\bm S.
\end{equation}
One is the rotational part $\bm R$, which is contributed to fluid rotation, and the other is non-rotational part $\bm S$, contributed to shear. This rotational part $\bm R$ is defined as rortex vector, and different kind of rortex vector in Cartesian coordinates were proposed \cite{liu-2,liu-3,liu-4,liu-5,liu-6,liu-7}.

Numerical simulations \cite{liu-2,liu-3,liu-4,liu-5,liu-6,liu-7} arguments that the rortex vector might be a promising and/or better quantity for vortex identification. If it is true, the question has become to find a general formulation of $\bm \omega=\bm \nabla \times \bm v= \bm R+\bm S$ which should be valid in any coordinate system.

Because the vorticity, $\bm \omega=\bm \nabla \times \bm v$, can be expressed in terms of velocity gradient $\bm \nabla  \bm v$, namely
\begin{equation}\label{q1}
  \bm \omega=\bm \nabla \times \bm v=\bm \varepsilon:\bm \nabla  \bm v,
\end{equation}
where the permutation tensor (symbol) $\bm \varepsilon=\varepsilon_{ijk}\bm e_i\bm e_j\bm e_k$, the base vector $\bm e_k$. This can be verified easily as follows: $\bm \varepsilon:\bm \nabla  \bm v=(\varepsilon_{ijk}\bm e_i\bm e_j\bm e_k):\bm e_p \nabla _p (v_q e_q)=\varepsilon_{ijk}(\nabla_pv_q) \bm e_i\bm e_j\bm e_k: \bm e_p \bm e_q=\varepsilon_{ijk}(\nabla_pv_q) \bm e_i(\bm e_j\cdot \bm e_p)(\bm e_k\cdot \bm e_q)=\varepsilon_{ijk}(\nabla_pv_q) \bm e_i\delta_{jp} \delta_{kq}=\varepsilon_{ipq}(\nabla_pv_q) \bm e_i=(\nabla_pv_q) \bm e_p\times \bm e_q=(\bm e_p\nabla_p) \times (v_q\bm e_q)=\bm \nabla\times \bm v$.

Therefore, to make the additive decomposition $\bm \omega=\bm \nabla \times \bm v= \bm R+\bm S$ possible, the velocity gradient $\bm \nabla  \bm v$ has to be decomposed additively as well, since the permutation tensor $\bm \varepsilon$ is a constant symbolic tensor. Otherwise, the decomposition $ \bm \omega=\bm \nabla \times \bm v=\bm \varepsilon:\bm \nabla  \bm v=\bm R+\bm S$ would not be compatible with the additive decomposition of the vorticity in Eq.\ref{liu}. It means that the decomposition of the vorticity $\bm \omega=\bm \nabla \times \bm v$ has become to a question of how to decompose the velocity gradient $\bm \nabla  \bm v$ additively.

Mathematically speaking, there is no unique additive decomposition of a tensor. The proper decomposition can only be defined if such decomposition works.

As the first additive decomposition, the velocity gradient $\bm \nabla  \bm v$ is decomposed to a symmetric part $\bm D=\bm D^T$ representing deformation rate and a skew-symmetric part $\bm W^T=-\bm W$ representing spin, namely
\begin{equation}\label{cauchy}
  \bm \nabla  \bm v=\bm D+\bm W,
\end{equation}
which is called the Cauchy-Stokes decomposition, and $\bm D = \frac{1}{2}[\bm \nabla  \bm v+(\bm \nabla  \bm v)^T]$ and $\bm W = \frac{1}{2}[\bm \nabla  \bm v-(\bm \nabla  \bm v)^T]$.

For the symmetric tensor $\bm D$, since $\bm \varepsilon:\bm D=(D_{23}-D_{32})\bm e_1+(D_{31}-D_{13})\bm e_2+(D_{12}-D_{21})\bm e_3=0$, hence the vorticity $\bm \omega$ in Eq.\ref{q1} can be expressed as follows
\begin{equation}\label{q2}
\begin{split}
  \bm \omega &=\bm \varepsilon:\bm D+\bm \varepsilon:\bm W=\bm \varepsilon:\bm W
  \end{split}
\end{equation}
From deformation geometry analysis, it has shown that the skew tensor $\bm W$ can only represent the infinitesimal rotation, where is the weak point of the Cauchy-Stokes decomposition that has been criticised \cite{liu-2,liu-3,liu-4,liu-5,liu-6,liu-7}.

The spin tensor $\bm W$ is not a proper SO(3) rotational tensor, because it does not satisfies the SO(3) definition:
\begin{equation}\label{w}
  \bm W\cdot \bm W^T=\bm W^T\cdot \bm W=-\bm W\cdot \bm W=-\bm W^2\neq \bm 1.
\end{equation}
It reveals that the spin $\bm W$ is not a proper candidate for finite rotation motion formulation. The SO(3) orthogonal rotation tensor $\bm Q$ is defined as following:
\begin{equation}\label{q3}
\begin{split}
  SO(3)=\{\bm Q: R^3 \rightarrow TR^3&|\bm Q\cdot \bm Q^T=\bm Q^T\cdot \bm Q=\bm 1,\det \bm Q=+1\},
   \end{split}
\end{equation}
To avoid the infinitesimal rotation, it would be an natural attempts to generalize the spin $\bm W$ to a rotation tensor $\bm Q$ that must be valid for any rotation instead of only for infinitesimal rotation.

In mathematics, given an antisymmetric tensor $\bm A$, its exponential map, ie. $\exp: \mathfrak{so}(3)\rightarrow SO(3); \bm A\mapsto e^{\bm A}=\sum_{k=0}^\infty \bm A^k/k!$, is always in SO(3), where $\mathfrak{so}(3)$ is Lie algebra of SO(3) and consists all skew-symmetric $3\times 3$ tensors. The proof uses the elementary properties of the tensor exponential
\begin{equation}\label{q4}
\begin{split}
  (e^{\bm A})^T \cdot e^{\bm A}&= e^{{\bm A}^T} \cdot e^{\bm A}=e^{-\bm A+\bm A}=e^{\bm A-\bm A}=e^{\bm A} \cdot (e^{\bm A})^T=\bm 1,
  \end{split}
\end{equation}
and $ \det (e^{\bm A})=e^{tr\bm A}=e^0=1$. In general, skew-symmetric tensors over the field of real numbers form the tangent space to the real orthogonal group $\displaystyle O(n)$ at the identity tensor; formally, the special orthogonal Lie algebra. In this sense, then, skew-symmetric matrices can be thought of as infinitesimal rotations. Another way of saying this is that the space of skew-symmetric tensors forms the Lie algebra $ \mathfrak{so}(n)$ of the Lie group $O(n)$.
The Lie bracket on this space is given by the commutator: $[\bm A,\bm B]=\bm A\cdot \bm B-\bm B\cdot \bm A$. It is easy to check that the commutator of two skew-symmetric tensors $\bm A$ and $\bm B$ is again skew-symmetric, ie., $[\bm A,\bm B]^T=(\bm A\cdot \bm B)^T-(\bm B\cdot \bm A)^T=(-\bm B)\cdot (-\bm A)-(-\bm A)\cdot (-\bm B)=\bm B\cdot \bm A-\bm A\cdot \bm B=-[\bm A,\bm B]$. The tensor exponential of a skew-symmetric tensor $\bm A$ is then an orthogonal tensor $\bm Q=e^{\bm A}=\sum_{k=0}^\infty \bm A^k/k!$.

Based on the above mathematical understanding, we know that the rotation tensor $\bm Q$ must be a tensorial function of an antisymmetric tensor. For the velocity gradient $\bm \nabla\bm v$, only antisymmetric tensor associated with the velocity gradient $\bm \nabla  \bm v$ is the spin tensor $\bm W$. Therefore, the rotation tensor $\bm Q$ should be an isotropic tensorial function of the spin tensor $\bm W$, namely $\bm Q=\bm f(\bm W)$. From the Cayley-Hamilton tensor representation theory \cite{truesdell-1,truesdell-2}, for the 3D spin tensor $\bm W$, the isotropic tensor function must be in following form:
\begin{equation}\label{w}
  \bm Q=\bm f(\bm W)=\alpha_0\bm 1+\alpha_1\bm W+\alpha_2 \bm W^2,
\end{equation}
where $\alpha_k,\,k=1,2,3$ can be determined later.

The mathematics suggests the rotation must be an exponential map: $e^{\beta \bm W}$, where $\beta$ is an arbitrary constant. If we set $\beta=1$, the simplest rotational tensor $\bm Q$ is hence proposed by the spin tensor $\bm W$ as follows
\begin{equation}\label{q5}
  \bm Q=e^{\bm W}=\exp \left\{\frac{1}{2}[\bm \nabla  \bm v-(\bm \nabla  \bm v)^T]\right\}.
\end{equation}
Hence we have an expression
\begin{equation}\label{q6}
  \bm Q=e^{\bm W}=\alpha_0\bm 1+\alpha_1\bm W+\alpha_2 \bm W^2.
\end{equation}
The eigenvalues of the spin tensor $\bm W$ are given by
\begin{equation}\label{q7}
  \lambda_0=0, \,\lambda_2=i\omega, \, \lambda_3=-i\omega,
\end{equation}
where $\omega=\sqrt{\bm \omega\cdot \bm \omega}$. Thus substituting the eigenvalues to Eq.\ref{q6}, we have
\begin{eqnarray}
  e^0 &=& \alpha_0 1+\alpha_1(0)+\alpha_2 (0)^2 \\
  e^{i\omega} &=& \alpha_0 1+\alpha_1(i\omega)+\alpha_2 (i\omega)^2 \\
  e^{-i\omega} &=& \alpha_0 1+\alpha_1(-i\omega)+\alpha_2 (-i\omega)^2
\end{eqnarray}
Solve the equations system, we obtain the coefficients
\begin{equation}\label{q8}
  \alpha_0=1,\, \alpha_1=\frac{\sin \omega}{\omega},\, \alpha_2=\frac{1-\cos \omega}{\omega^2}.
\end{equation}
The rotation tensor in Eq.\ref{q5} then can be constructed in terms of spin tensor $\bm W$ as follows
\begin{equation}\label{q9}
\begin{split}
  \bm Q&=\bm 1+\frac{\sin \omega}{\omega} \bm W+\frac{1-\cos \omega}{\omega^2} \bm W^2\\
  &=\bm 1+\frac{\sin \omega}{2\omega} [\bm \nabla  \bm v-(\bm \nabla  \bm v)^T]+\frac{1-\cos \omega}{4\omega^2}[\bm \nabla  \bm v-(\bm \nabla  \bm v)^T]^2.
  \end{split}
\end{equation}
With the rotation tensor $\bm Q$ defined in Eq.\ref{q9}, we can propose an new additive decomposition of the velocity gradient $\bm \nabla  \bm v$ as follows
\begin{equation}\label{q10}
  \bm \nabla  \bm v=\bm K+\bm Q,
\end{equation}
and $\bm K= \bm \nabla  \bm v-\bm Q$, which can be expressed in terms of $\bm D$ and $\bm W$, namely
\begin{equation}\label{p3}
  \bm K=\bm D-\bm 1-\frac{1-\cos \omega}{\omega^2} \bm W^2+(1-\frac{\sin \omega}{\omega}) \bm W.
\end{equation}
Therefore, together with Eqs \ref{q9},\ref{q10} and \ref{p3}, the velocity gradient$\bm \nabla \bm v$ has been successfully split into two parts. The additive decomposition in Eq.\ref{q10} is valid for any rotation due to the introduction of rotation tensor $\bm Q$. We have done the task to decompose the velocity gradient into two parts, one must be a rotation tensor.

As a consequence of Eq.\ref{q10}, if set $\bm K$ as a symmetric tensor, the skew-symmetric part of the $\bm K$ must be vanish, namely $Skew(\bm K)=(1-\frac{\sin \omega}{\omega})\bm W=\bm 0$, we will have a special additive decomposition as follows
\begin{eqnarray}\label{b}
  \bm K &=\bm D-\bm 1-\frac{1}{1+\cos \omega} \bm W^2 \\
  \bm Q &=\bm 1+\bm W+\frac{1}{1+\cos \omega} \bm W^2
\end{eqnarray}
Let's go back to the vorticity by substituting the new additive decomposition Eq.\ref{q10} into Eq.\ref{q1}, the vorticity is hence expressed in terms of $\bm K,\, \bm Q$
 \begin{equation}\label{q12}  \bm \omega =\bm \varepsilon:(\bm K+\bm Q).\end{equation}
Although we have the new additive decomposition $\bm \nabla \bm v=\bm K+\bm Q$, due to the mathematical nature of the permutation tensor, antisymmetric under the interchange of any two slots \cite{levi}, The permutation tensor $\bm \varepsilon$ will cancel out the contribution from any symmetric part of the velocity gradient by the double dot product ":".

This can be easily proved by substituting the expressions of both $\bm K$ and $\bm Q$ into Eq.\ref{q12}.
\begin{equation}\label{q13}
\begin{split}
  \bm \omega&=\bm \varepsilon:[\underbrace{\bm D-\bm 1-\frac{1-\cos \omega}{\omega^2} \bm W^2+(1-\frac{\sin \omega}{\omega}) \bm W}_{=\bm K}]
  +\bm \varepsilon:[\underbrace{\bm 1+\frac{\sin \omega}{\omega} \bm W+\frac{1-\cos \omega}{\omega^2} \bm W^2}_{=\bm Q}].
\end{split}
\end{equation}
Notice $\bm \varepsilon: \bm 1=0$, $\bm \varepsilon:(\bm D-\bm 1)=0$ and $\bm \varepsilon:\bm W^2=0$ due to the symmetric nature of tensors of $\bm 1,\, \bm D$ and $\bm W^2$ \footnote{Let $\bm Z=\bm W^2$, then $\bm Z^T=(\bm W\cdot \bm W)^T=\bm W^T\cdot (\bm W^T)=(-\bm W)\cdot (-\bm W)=\bm W\cdot \bm W=\bm W^2$, hence $\bm Z^T=\bm Z$.}, Eq.\ref{q13} is reduced back to Eq.\ref{q2}, i.e., $
  \bm \omega =\bm \varepsilon:(1-\frac{\sin \omega}{\omega}) \bm W+\bm \varepsilon:(\frac{\sin \omega}{\omega} \bm W)=\bm \varepsilon:\bm W$, which reveals that the vorticity $\bm \omega$ is only affected by the antisymmetric part of the velocity gradient $\bm \nabla \bm v$ and
  has nothing to do with its symmetric part $\bm D$.

In summary, we have successfully proposed an new additive decomposition $\bm \nabla \bm v=\bm K+\bm Q$ that is different from Cauchy-Stokes decomposition. The new decomposition might provided a new way of thinking on the vortex vector, however, what is the relationship between the vortex vectors in \cite{liu-2,liu-3,liu-4,liu-5,liu-6,liu-7} and the decomposition in Eq.\ref{q10} is still an open question and needed for further investigation.

\textbf{Acknowledgements}: It is my great pleasure to have shared and discussed some of the above with Michael Sun from Bishops Diocesan College, whose pure and direct scientific sense inspired me.

\nolinenumbers

%
%
%

\end{document}